\documentclass[a4paper]{jpconf}

\usepackage{natbib}    % for building bibliography with \cite{}
\usepackage{graphicx}  % for including PostScript figures

\newcommand{\numu}{\mbox{$\nu_{\mu}$}\hspace{1mm}}                   % nu_mu
\newcommand{\nue}{\mbox{$\nu_{e}$}\hspace{1mm}}                      % nu_e
\newcommand{\nutau}{\mbox{$\nu_{\tau}$}\hspace{1mm}}                 % nu_tau
\newcommand{\numubar}{\mbox{$\overline{\nu}_{\mu}$}\hspace{1mm}}     % nu_mubar
\newcommand{\nuebar}{\mbox{$\overline{\nu}_{e}$}\hspace{1mm}}        % nu_ebar
\newcommand{\dmsq}{\mbox{$| \Delta m^2_{32} |$}\hspace{1mm}}
\newcommand{\sinsq}{\mbox{$\sin^2(2 \theta_{23})$}\hspace{1mm}}
\begin{document}

\title{New Results from the MINOS Experiment}

\author{Hugh Gallagher for the MINOS Collaboration}
\ead{hugh.gallagher@tufts.edu}
%%%%%%%%%%%%%%%% ADDRESSES

\address {Physics Department, Tufts University, Medford, MA, 02155, USA}
%%%%%%%%%%%%%%%% ABSTRACT

\begin{abstract}
In this paper we present the latest results from the MINOS Experiment.  This includes
 a new measurement of the oscillation parameters 
($| \Delta m^2_{32} |$,  $\sin^2(2 \theta_{23})$) based on 
 $3.36 \times 10^{20}$ protons-on-target of data and 
a first analysis of neutral current events in the Far Detector.  The prospects for $\nu_e$ 
appearance measurements in MINOS are also discussed.  
\end{abstract}

\section{Introduction}
MINOS is a two detector long-baseline experiment utilizing the Neutrinos 
at the Main Injector (NuMI) beamline at Fermilab.  
The Near and Far 
Detectors have masses of 0.98 kton and 5.4 kton with the Far Detector 
located at a distance of 735 km in the Soudan Mine in Minnesota.     
The experiment is designed to explore oscillations at the large
$\Delta m^2$ previously probed by atmospheric neutrino experiments and the K2K 
experiment \cite{Ahn:2006zza,Ashie:2004mr,Ashie:2005ik}.           
The experimental goals include precision measurements of the oscillation 
parameters, testing for exotic neutrino disappearance explanations, 
and searching for subdominant $\nu_e$ appearance.  
In addition, charge-separated 
atmospheric neutrino and cosmic ray studies can be carried out with the Far
Detector and neutrino cross section measurements can be carried out in the 
high-rate environment of the Near Detector.  
The published result from the experiment on the measurement of 
oscillation parameters was
based on an exposure of $1.27 \times 10^{20}$ protons-on-target (POT) \cite{Michael:2006rx}.

In the NuMI beam 120 GeV protons from the Fermilab Main Injector are extracted 
in a 10 $\mu$s spill every 2.2 s and strike a water-cooled graphite 
target, producing kaons and pions.  Two parabolic magnetic horns focus charged  
secondaries, producing a beam that is 92.9\% \numu{}, 
5.8\% \numubar{} and 1.3\% \nue{}+\nuebar{}.  
The beam is now in its third
year of operation and is obtaining intensities of $3.0\times 10^{13}$ POT/spill for a beam power 
of 0.275 MW, and is regularly achieving $10^{18}$ POT/day.   The energy spectrum of the 
neutrino beam can be changed by moving the position of the target relative to the horns. 
Data taken in several different target configurations are used to improve modeling of hadron 
production off the target and thereby improve the prediction of the neutrino 
beam \cite{Adamson:2007gu}.

\section{MINOS Analyses}

The analyses presented here have several steps in common and are labeled by the 
dominant scattering mechanism.  In the `CC Analysis' we compare selected \numu{} 
charged current interactions 
in the Near and Far Detectors to make a measurement of the oscillation parameters 
\dmsq{} and $\sin^2(2\theta_{23})$.  
In the `NC Analysis' we compare selected neutral current events in the two 
detectors to investigate the possibility of oscillations into sterile neutrinos.  In the `\nue{} Analysis' we compare $\nu_e$-selected events in the two detectors to search for subdominant oscillations and 
non-zero $\theta_{13}$.    All have been or are being conducted as `blind analyses', i.e. 
the full set of Far Detector data sensitive to the phenomena in 
question is not examined until the complete analysis procedure is defined.  For all analyses
the event selection is applied first to the high-statistics data set from the Near Detector.  
This measurement is 
then used to predict what will be measured in the Far Detector 
under a particular oscillation scenario \cite{Adamson:2007gu}.
 
The data are divided into three data sets, Run 1, Run 2, and Run 3, 
separated by Fermilab shutdowns.  The Run 1 data were analyzed in our 2006 publication 
\cite{Michael:2006rx}.    
The CC analysis 
presented here includes all Run 1 and Run 2 data including 
$1.5 \times 10^{19}$ POT of data in a high-energy configuration 
that were taken at the start of Run 2.  
The NC analysis presented here uses the Run 1 data and a subset of the Run 2 
data.

\subsection{Charged Current Analysis}
The measurement of oscillation parameters is made through a comparison of charged
current muon neutrino interactions in the two detectors. 
Charged current (CC) vs. neutral current (NC)  event classification is
performed with a $k$-nearest neighbor-based 
algorithm with four inputs: track length (in planes),
and for hits on the track the
mean pulse height,
fluctuation in pulse height, and
transverse track profile \cite{Ospanov:Thesis}.  Relative to the previous
analysis this new classifier, in conjunction with new event reconstruction,
improves the efficiency for selection of \numu{} CC events in the absence of oscillations 
from 75.3\% to 81.5\% and reduces the NC contamination from 1.8\% to 0.6\%.

%A measured set of CC events in the Near Detector is then used to predict what 
%ill be seen in the Far detector.  
%Although the energy spectra of neutrinos at the two detectors are different, they 
%come from the same parent pions.  Imagine taking a particular bin in true neutrino 
%energy in the Near detector, and tracking these back to their parents in the beam 
%pipe, and then asking what neutrino energies these would give in the Far detector.  
%This relationship can be encoded in a matrix which encompasses the geometry of the 
%beamline and the kinematics of pion/kaon decay.  The full MC is used to provide
%smearing and acceptance corrections to the data to get to a true neutrino 
%distribution, which is then extrapolated to the far detector using the matrix, 
%and energy smearing corrections and acceptance are again applied.

\begin{figure} 
\center
\includegraphics[width=20pc]{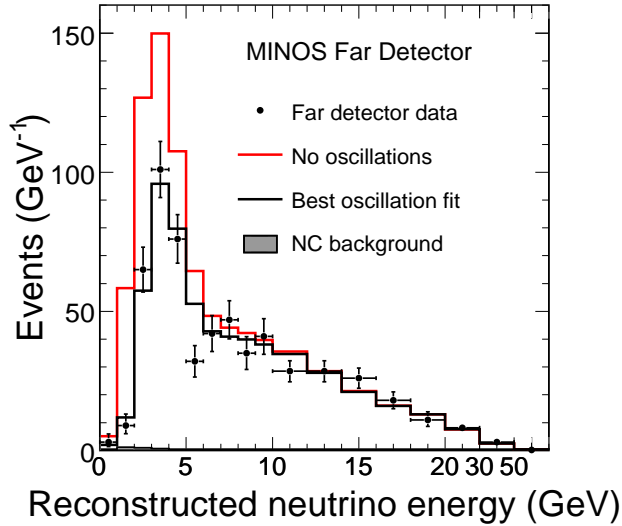}
\caption{
Energy spectrum of CC events in the Far Detector compared with the Monte Carlo
expectation with no oscillations (red) and with the best fit oscillation 
parameters (black).   
}           
\label{fig:ColourSpectrum}
\end{figure}

848 events are observed in the Far Detector, while $1065\pm60$ (syst.) are expected 
in the absence of oscillations.   
The energy spectrum of events in the Far Detector is shown in Figure 
\ref{fig:ColourSpectrum}.
The impact of different sources of systematic uncertainty were evaluated by fitting modified MC in place of the data, and the three largest were included
as nuisance parameters in the oscillation fit.  These are the relative
normalization of the Near and Far detectors, the overall hadronic energy
scale, and the NC background which are included with penalty terms of
4\%, 10.3\%, and 50\%, respectively.
Expected backgrounds in the CC sample are  5.9 NC events, 1.5 \nutau{} CC events, 2.3 events coming
from \numu{} CC interactions in the rock surrounding the detector cavern,
and 0.7 events from cosmic rays.  

%The oscillation fits include the following:
%\begin{itemize}
%\item
%    Run I LE data:  $1.27 \times 10^{20}$ POT
%\item
%    Run II LE data: $1.94 \times 10^{20}$ POT
%\item
%    Run II HE data:  $0.15 \times 10^{20}$ POT
%\end{itemize}

\begin{figure}
\center
\includegraphics[width=20pc]{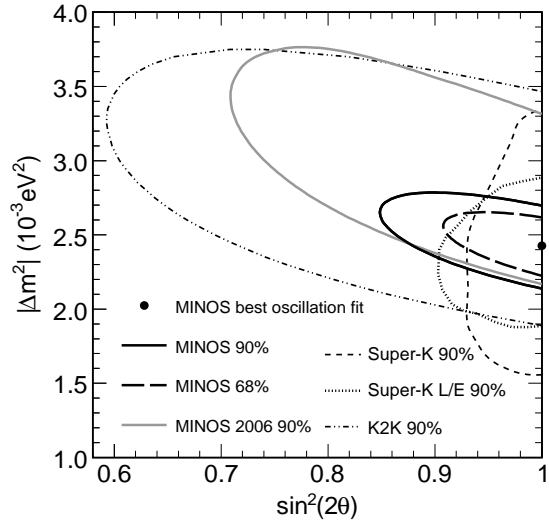}
\caption{Allowed region in \dmsq{}, \sinsq{} parameter space including systematic errors. Also shown 
are the allowed regions from the K2K and Super-K analyses \cite{Ahn:2006zza,Ashie:2004mr,Ashie:2005ik}
and our earlier result \cite{Michael:2006rx}.}           
\label{fig:PRLContour}
\end{figure}

The data are fit using the standard two-generation oscillation formula with \sinsq{}
constrained to be in the physical region.
% and a $\chi^2$ statistic that incorporates Poisson
% errors and systematic errors:
We obtain \dmsq{}$=(2.43\pm0.13)\times10^{-3}$ eV$^2$ and $\sin^2(2\theta_{23})>0.95$ at 68\% CL.   
At the best fit point the 
value of $\chi^2$ is 90 for 97 degrees of freedom.    Figure \ref{fig:PRLContour} shows the 
allowed region from this fit compared with the results of previous experiments.   
If the fit is allowed to go into the unphysical region the best fit value moves to
$\Delta m^2=2.33\times10^{-3}$ eV$^2$ and $\sin^2(2\theta)=1.07$ with a reduction in 
$\chi^2$ of 0.6.   We have carried out a Monte Carlo study in which the input true value was 
that obtained from our constrained fit, and found that 26.5\% of unconstrained fits have a 
best fit with a value of $\sin^2(2\theta)\ge1.07$. 

%\begin{equation}
%\chi^2 = \sum_{nbins}(2(e_i-o_i)+2o_i \ln(o_i/e_i)) + \sum_{nsys} \frac{\Delta s_j^2}{\sigma_{s_j}^2}.
%\end{equation}

\begin{figure}
\center 
\includegraphics[width=20pc]{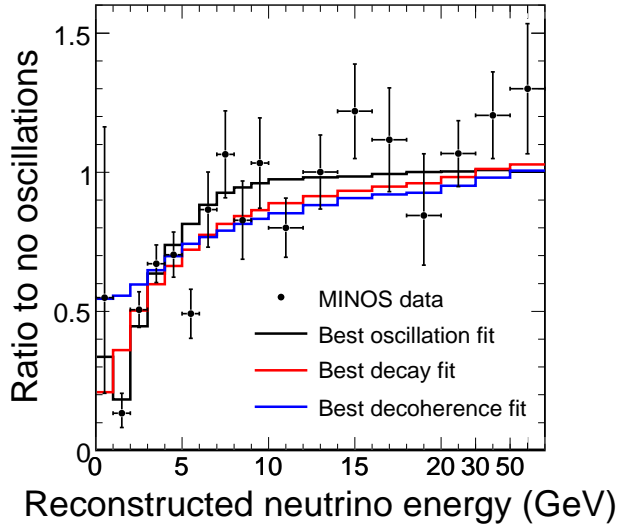}
\caption{Best oscillation fit compared with the best fits to neutrino decay (color) and 
decoherence (color).}           
\label{fig:ColourRatio}
\end{figure}

We have also explored how well this data is described by phenomena other than oscillations. 
In particular we have fit the data to models for neutrino decay \cite{Barger:1998xk} 
and decoherence \cite{Fogli:2003th}. 
The ratio of the Far Detector data to no oscillations is shown in Figure \ref{fig:ColourRatio}
compared with the results from the best fits to oscillations, decay, and decoherence. 
The $\chi^2$ values for the decay and decoherence models are 104 and 123, they are thus disfavored
relative to the oscillation hypothesis by 3.7$\sigma$ and 5.7$\sigma$, respectively.

\subsection{Neutral Current Analysis}

Oscillations into sterile neutrinos would affect the number of neutral current
interactions in the Far detector.
The goal 
of the analysis is to look for NC disappearance at the Far detector, and in particular 
an energy dependent depletion of NC events as evidence for oscillations into sterile 
neutrinos.  
The analysis described here is based on 
$2.46 \times 10^{20}$ POT of data in the Far Detector. 

Pre-selection cuts are made for fiducial volume, beam quality, and to remove
overlapping events in the high-rate environment of the Near Detector.   
An event in either detector is then 
classified as NC if it has a reconstructed shower, is shorter than 60 planes, 
and does not have a track extending more than 5 planes beyond the end of the shower. 
With these cuts NC events are selected with 90\% efficiency and 60\% purity.  
More information about the NC analysis can be found in Ref. \cite{Adamson:2008jh}.

In our first examination of NC events in the Far Detector we would like to be as 
model-independent as possible.  To this end we compare our observations with the 
predictions of the standard picture of oscillations between three active flavors.    
The NC measurement in the Near Detector is 
extrapolated to the Far Detector assuming the previous MINOS
CC best fit result for the oscillation parameters of $|\Delta m^2_{32}|=2.38 \times 10^{-3}$ eV$^2$, 
$\sin^2(2\theta_{23})=1$ \cite{Collaboration:2007zza}.  In comparing the 
Far Detector NC energy spectrum with the expectation of standard 3-flavor oscillation physics, 
one also has to assume some values for the other oscillation parameters.  For this analysis 
$\Delta m^2_{21}=7.59 \times 10^{-5}$ eV$^2$, and $\theta_{12}=0.61$ are taken from 
combined KamLAND+SNO fits \cite{Abe:2008ee}.  Since CC $\nu_e$ events will also be classified
as NC by this analysis the NC rate in the Far Detector will also be affected by sub-dominant
$\numu \rightarrow \nu_e$ mixing.  The data are compared separately to $\theta_{13}=0$ and 
$\theta_{13}=0.21$.  

\begin{figure}
\center
\includegraphics[width=25pc]{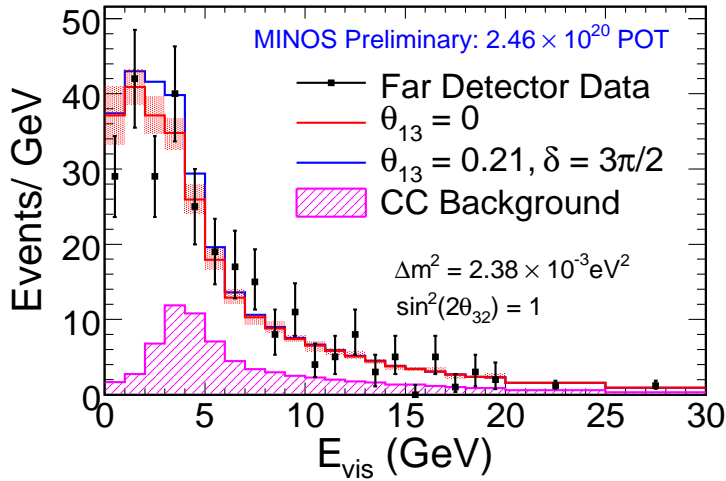}
\caption{Reconstructed energy spectrum of NC-selected events in the Far Detector.  
Data are black points with error bars, blue and red curves are the Monte Carlo expectation
including systematic uncertainties for 
two different values of $\theta_{13}$.   The hatched distribution shows the expected 
$\nu_\mu$ CC background. 
}           
\label{fig:33FlavorPrediction_NC}
\end{figure}

The spectrum of NC events in the Far Detector compared with the expectation under these two 
oscillation scenarios is shown in Figure \ref{fig:33FlavorPrediction_NC}, together with the 
contribution to the sample from true CC interactions.    
The basic statistic for comparison is 
the number of events in energy ranges 0-3 GeV, 0-5 GeV, and over all energies.   
Table \ref{tab:nc} shows the results of these event rate comparisons for the 
$\theta_{13}=0$ case, the stated significances are slightly larger for $\theta_{13}$ at 
the Chooz limit.   From these data we find that for NC interactions with visible 
energy less than 3 GeV the fraction that disappear is less than 35\% at 90\% CL, and 
that over all energies the fraction that disappear is less than 17\% at 90\% CL. 

\begin{table}[htb]
\begin{center}
\begin{tabular}{|c|r|r|r|}
\hline
Energy Range  &       Data      &    Expectation      &  Significance ($\sigma$) \\
\hline
0-3 GeV       &        100      &   115.16 $\pm$7.67  &     1.15                 \\
0-5 GeV       &        165      &   175.92 $\pm$10.42 &     0.65                 \\
0-120 GeV     &        291      &   292.63 $\pm$15.02 &     0.10                 \\
\hline
\end{tabular}
\end{center}
\caption{Observed and expected NC events in the Far Detector for $\theta_{13}=0$.}
\label{tab:nc}
\end{table}

\subsection{Electron Neutrino Appearance}

Electromagnetic showers can be identified in MINOS by the 
characteristic topology of their energy deposition.  The challenge in MINOS
is separating these events from the much more numerous NC interactions which 
can have similar topologies.  \nue{} identification in both detectors 
is performed with an artificial neural network.  
Because of difficulties in modeling hadronic
showers at NuMI/MINOS energies there is a large uncertainty on the prediction 
of the $\nu_e$-selected sample in the Near Detector \cite{Yang:2007zzt} and the observed rate 
disagrees with the Monte Carlo expectation by around 20\%.  

The measurement of Near Detector $\nu_e$-selected events cannot be immediately extrapolated 
to the Far Detector since the sample is composed of both \numu{} NC and CC events, which differ
in their extrapolation due to oscillations.    Two methods
have been developed to independently measure the CC and NC contributions
to \nue{} backgrounds in the Near Detector.  These independent, data-driven
methods are in good agreement regarding the NC and CC \numu{} backgrounds.
The first technique uses the identified \numu{} CC sample,
removes the muon, and analyzes the remaining hadronic shower as if it
were a NC event.  The second technique uses the fact that the data set obtained
when the magnetic horn is not powered has a very different mix of NC/CC \numu{} events in the
$\nu_e$-selected Near Detector sample.  The so-called `horn-on/horn-off' method takes advantage
of this fact to deconvolute the NC and CC contributions to the measured Near 
Detector sample.  

\begin{figure}[htb]
\center
\includegraphics[width=25pc]{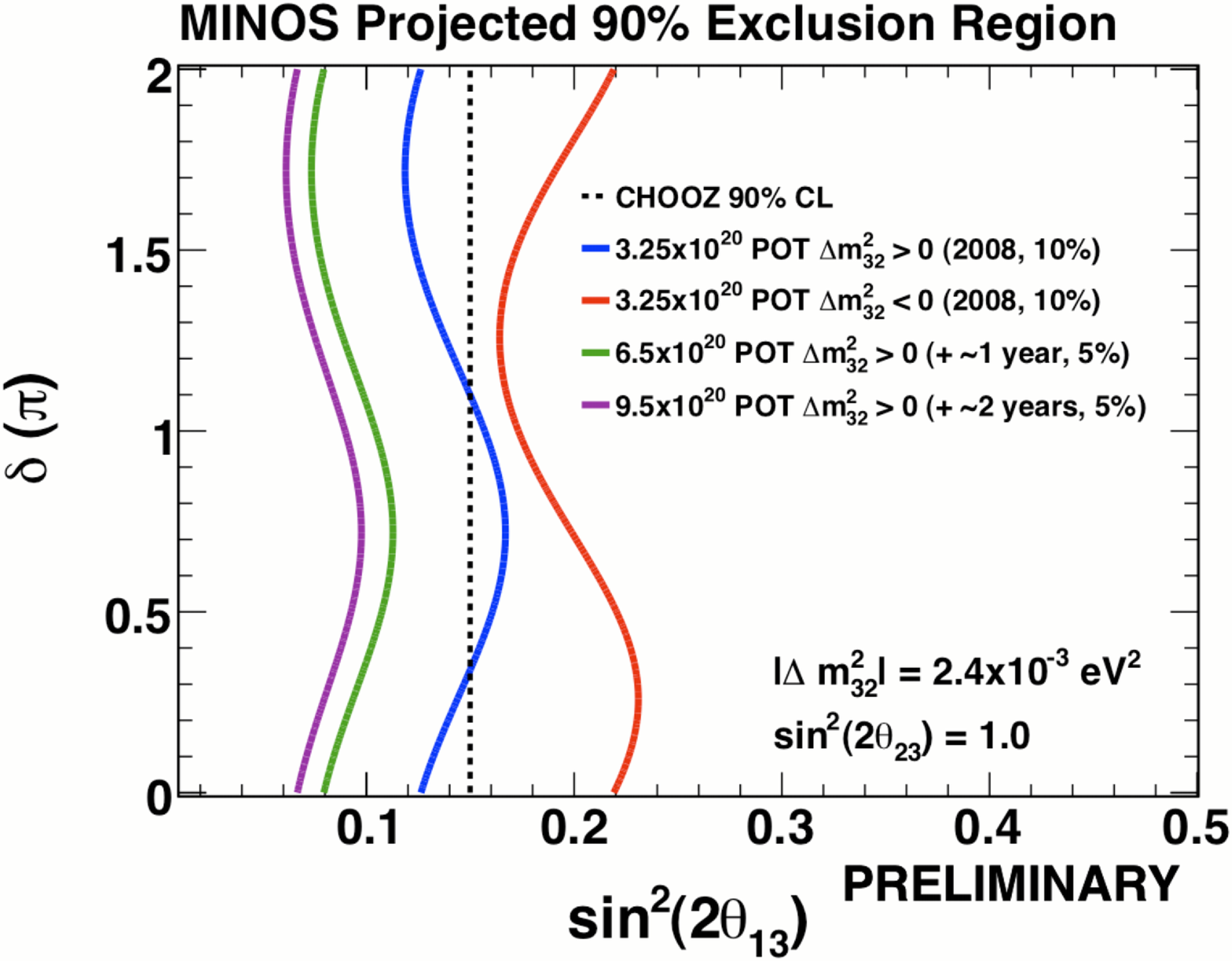}
\caption{Sensitivity to electron neutrino appearance for three different 
exposures including 10\% systematic errors.}
\label{fig:FutureLin_spreaddm24}
\end{figure}

With these measurements in the Near Detector we can now calculate more accurately
our expected sensitivity to \nue{} appearance in a Far Detector measurement.
Figure \ref{fig:FutureLin_spreaddm24} shows the MINOS projected
90\% CL exclusion region.  At the CHOOZ limit we expect 12 \nue{}
signal events and 42 background events with our current exposure of
$3.25\times10^{20}$ POT.  The reliability of the \nue{}
appearance algorithm as well as the extrapolation technique will
be demonstrated on several sideband regions in the Far Detector
which are expected to have limited
sensitivity to \nue{} appearance.

\section{Acknowledgements}
This work was supported by the US DOE and NSF; the UK STFC; 
the State and University of Minnesota; the University of Athens, Greece;
and Brazil's FAPESP and CNPq.  We are grateful to the Minnesota Department
of Natural Resources, the crew of the Soudan University Laboratory, and the 
staff of Fermilab for their contribution to this effort. 
The author was supported by DOE grant DE-FG02-92ER40702 and would also 
like to acknowledge the excellent work by 
the organizers of the Neutrino 08 Conference.
% BIBLIOGRAPHY
%------------------------------------------------------------------------------
\bibliographystyle{iopart-num}
\bibliography{minos_nu08}

\end{document}